# Intégration du contrôle automatique dans la maîtrise statistique des procédés

**Wafik Hachicha\*,\*\* — Faouzi Masmoudi\*,\*\*\***

**Ahmed Ammari\*\* — Sami Abidi\*\***

\* *Unité de recherche de Mécanique, Modélisation et Production (U2MP)*
 *Ecole Nationale d'ingénieurs de Sfax,*
*B.P. W, 3038 Sfax, Tunisie.*
*wafik_hachicha@yahoo.fr*

\*\* *Institut Supérieur de Gestion Industrielle de Sfax (ISGI)*
*Route El - Meharza km 1,5 ; B.P. 954, 3018 Sfax, Tunisie*

\*\*\* *Dep. de Génie Mécanique, Ecole Nationale d'ingénieurs de Sfax,*
*B.P. W, 3038 Sfax, Tunisie.*
*faouzi.masmoudi@enis.rnu.tn*

RÉSUMÉ. *La Maîtrise Statistique des Procédés (MSP) et le Contrôle Automatique des Procédés (CAP) poursuivent un même objectif : réduire la variabilité du procédé et le garder sur la cible. Ce travail porte sur la proposition d'un modèle d'intégration du CAP dans la MSP qui se base sur la discrétisation des fonctions de transfert relatives à chaque composante du procédé. Nous avons proposé d'une part, une nouvelle règle de contrôle qui se base sur un système de premier ordre. D'autre part, nous avons montré comment établir des cartes de contrôle à un procédé de type AR (1). A l'aide d'expériences de simulation, nous avons montré que la règle que nous avons proposée a réduit la variabilité en la comparant à celle proposée en littérature.*

ABSTRACT. *The Statistical Process Control (SPC) and the Automated Process Control (APC) have a common goal: achieve optimal product quality by controlling variations in the process. The work in this paper will present a developed integration methodology of the APC in the SPC which is based on discretization of the transfer functions relating to each component of the process. We proposed on the one hand, a new control rule which is based on a system of first order. In the other hand, we showed how to establish control charts to a process of the type AR (1). Using simulation experiments, we showed that the proposed control rule reduced variability by comparing it with that proposed in literature.*

MOTS-CLÉS : *maîtrise statistique des procédés (MSP), contrôle automatique des procédés (CAP), modèle AR (1), discrétisation, simulation, variabilité.*

# 1. Introduction

La recherche de la qualité est devenue un point-clé de la compétition du fait de l'importance de l'offre par rapport à la demande. Ainsi, l'obtention de la qualité des produits passe le plus souvent par la mise en place d'un système d'assurance qualité et par l'utilisation des outils et des méthodes de la qualité tant au niveau de la conception que de la réalisation des produits. La statistique a une place essentielle dans ce mouvement qualité par les techniques et la méthodologie qu'elle propose pour aider à une résolution objective, méthodique et rigoureuse des problèmes (Hubérac, 2001).

La Maîtrise Statistique des Procédés (MSP) est une discipline qui s'inscrit dans une stratégie de prévention pour améliorer la qualité d'une production. Elle comporte une suite d'analyses comme une réflexion sur le processus, l'identification des caractéristiques significatives du processus et du produit, la validation de l'outil de production et de son aptitude à fournir ce qu'on attend de lui, et enfin la mise en place de cartes de contrôle (Pillet, 2005), (Halais, 2002). L'objectif de l'établissement des cartes de contrôle est de vérifier à chaque instant si un procédé est « sous contrôle statistique » c'est-à-dire subit uniquement des causes de variation dites « normales ».

Une seconde discipline : le contrôle automatique des procédés (CAP) fournit des techniques, régulation et autres, pour ajuster un procédé dans le but de le garder le plus stable possible. Le CAP tente à chaque instant de garder les variables qui caractérisent le produit sur leur cible, il est appliqué spécialement dans des procédés dynamiques où, sans contrôle, les caractéristiques du produit ne restent pas naturellement stables même quand le procédé est en parfait état de marche (Ostertag, 2004).

Les adeptes du CAP reprochent à ceux du MSP que les cartes de contrôle sont totalement inefficaces pour conduire un procédé. Chacune de ces remarques a été discutées avec plus de nuances par différents auteurs (Deming, 1986), (MacGregor, 1990) et (Box et Kramer, 1992).

Durant ces dernières années les spécialistes se sont rendus compte qu'au lieu de tenter de travailler dans un même but avec deux disciplines parallèles, il pourrait être plus constructif d'intégrer ces deux types de techniques en retenant les avantages de chacune d'elles. Le rôle du MSP est de détecter les dérives ou les perturbations que peut subir un procédé et de l'améliorer en supprimant ces causes de variation (Ryan, 1989). Le rôle de CAP est d'ajuster en continu le procédé pour l'empêcher de dévier de sa cible mais pas de supprimer les causes de variation. La MSP et le CAP poursuivent donc tous les deux un même objectif : réduire la variabilité du procédé et le garder sur la cible mais proposent de l'atteindre par des méthodes différentes (Vander Wiel et al., 1992) et (Tucker et al., 1993) introduisent une technique appelée ASPC (Algorithmic Statistical Process Control) qui combine les deux approches et l'appliquent dans l'industrie des polymères. (Montgomery et al., 1994) montrent que la conciliation de ces techniques peut globalement apporter des réductions de variabilité d'un procédé que chacune des disciplines séparée ne peut pas atteindre seule. (Schippers, 2001) met l'accent sur l'importance de l'intégration du maintenance totale productive avec les techniques MSP et CAP pour maîtriser les processus. (Venkatesan, 2003) présente les avantages et les inconvénients de l'intégration entre MSP et CAP qui dépendent fortement du nature du bruit de l'environnement.

Le premier objectif de cet article est de proposer la structure d'un modèle d'intégration du CAP dans la MSP. L'exposé de la démarche se base sur un exemple présenté dans (Fearn et Maris, 1990) qui consiste à un procédé de mélange de protéines en poudre à de la farine brute visant à obtenir une farine qui contient un taux de protéines le plus constant possible autour d'une valeur cible. Le deuxième objectif est de réduire la variabilité du processus en définissant une autre règle de contrôle plus efficace que celle proposée en littérature.

Dans un premier temps, nous nous intéressons à définir le procédé d'addition de gluten dans un moulin à farine. Par la suite, nous proposons la structure du modèle d'intégration proposé, ce paragraphe présentera d'une part la règle de contrôle proposée par (Fearn, 1990) et celle que nous avons proposée. La dernière partie s'intéresse à la présentation des résultats de simulation de chaque règle de contrôle. Enfin, nous concluons sur des perspectives d'évolution de notre travail.

## 2. Présentation du procédé

### 2.1. *Enoncé du problème*

Le procédé, sujet de cette étude, est celui présenté par (Fearn et Maris, 1990). Il consiste à un procédé d'ajout de protéine à une farine brute. La farine de blé moulue en industrie ne contient en général pas suffisamment de protéines pour permettre de fabriquer du pain. Une solution adoptée consiste à ajouter à la farine moulue des protéines supplémentaires sous forme de poudre de gluten séché. Typiquement, la farine de départ peut contenir 10 % de protéines et le contenu désiré valoir 15,1 %. Il faudra dans ce cas ajouter environ 6 g de gluten par 100g de farine pour atteindre cette valeur. Dans le procédé étudié, le gluten est ajouté à la fin du procédé de broyage par l'intermédiaire d'un distributeur dont il est possible de régler le débit.

Il est important que le contenu en protéines de la farine produite ne soit pas plus faible que la valeur cible pour ne pas sortir des spécifications. Il est également important de ne pas mettre trop de gluten car ce produit est coûteux. De plus, la farine brute, n'a pas nécessairement un contenu de protéines constant dans le temps. Ces différents éléments montrent la nécessité d'avoir une règle qui permette de choisir la quantité optimale de gluten à introduire à chaque instant. La figure 1 présente le procédé d'addition de gluten et le système disponible pour contrôler la quantité de gluten à introduire dans le procédé. La farine brute arrive dans le système après broyage à un débit constant connu (100 g/sec par exemple) et le gluten y est ajouté à un certain débit variable contrôlé par le système. Le taux exact en protéines de la farine de départ est inconnu. La farine "améliorée" ou traitée passe ensuite dans un système de mesure automatique qui permet de prélever chaque minute une certaine quantité de farine et d'en mesurer le contenu en protéines. Le résultat de l'analyse est ensuite envoyé à une unité de contrôle qui va comparer le taux mesuré à la valeur cible et transmettre si nécessaire un ajustement à faire au distributeur de gluten pour garder la quantité de protéines de la farine traitée la plus proche possible de la valeur cible.

Les notations de la figure 1, utilisées pour décrire les différentes variables du système sont :

$\tau$ : Valeur cible du taux en protéines de la farine traitée sortant du système. Ce taux peut varier en fonction du type de la farine produite.

$i$ : Indice de l'instant où des prélèvements de farine traitée, des mesures du taux de protéines et des ajustements du débit de gluten sont faits. L'unité de temps est la seconde.

$X_i$ : Taux de protéines dans la farine brute arrivant dans le système à l'instant i (en %). La valeur exacte de ce taux est inconnue et variable. Supposons ici que le taux moyen est de 10 %.

D : Débit de farine arrivant dans le système. D est fixé, ici, à 100g par seconde.

$u_{i-1}$ : Débit de gluten entrant dans le système à l'instant i (en g/sec). L'indice (i-1) est utilisé car ce débit est choisi sur base des mesures faites à l'instant (i-1).

$u_0$ : Débit de gluten à l'instant 0 (début de production). Il sera fixé ici à 6g/sec.

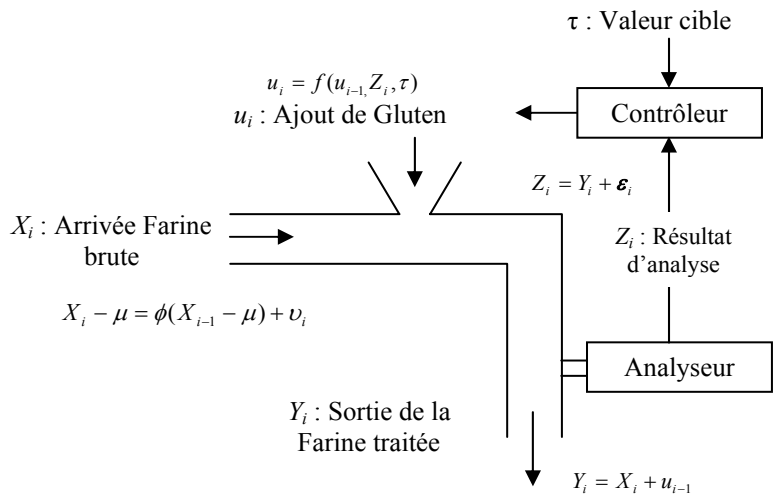

**Figure 1.** *Procédé de mélange de gluten à la farine (Fearn et Marris, 1990)*

$Y_i$ : Taux de protéines réel dans la farine sortant du système à l'instant i. Ce taux est mesuré par l'intermédiaire du système d'analyse.

$Z_i$ : Taux de protéines dans la farine sortant du système mesuré à l'instant i.

$\tau_0$ : Taux de protéines dans la farine traitée à l'instant initial.

**2.2. *Modèles statistiques réagissant le système***

Les équations suivantes peuvent être écrites pour décrire les relations existant entre les différentes variables du système et les perturbations aléatoires qui peuvent les affecter.

Le taux en protéines de la farine à la sortie est lié au taux à l'entrée et au débit de gluten par l'équation de conservation de masse suivante :

$$Y_i = \frac{(u_{i-1} + X_i)*100}{100 + u_{i-1}} \qquad [1]$$

Il est à noter que

$$\frac{100}{100 + u_{i-1}} \approx \frac{100}{100 + u_0} = cste \qquad [2]$$

Pour simplifier la manipulation des variables dans la suite nous allons redéfinir *τ, Y et Z* comme des quantités de protéines se trouvant dans de farine à la sortie. Les nouvelles variables sont simplement les anciennes multipliées par la constante de valeur 1,06.

L'approximation de l'équation [2] permet d'écrire l'équation [1] sous la forme simplifiée suivante :

$$Y_i = u_{i-1} + X_i \qquad [3]$$

Deux types de perturbations aléatoires peuvent affecter le système :

– Le taux de protéines dans la farine brute peut varier dans le temps autour d'une valeur moyenne *μ*. Nous supposerons que ce taux suit le modèle AR (1) suivant :

$$X_i - \mu = \phi(X_{i-1} - \mu) + \upsilon_i \text{ Où } X_0 = \mu \qquad [4]$$

Avec $\upsilon_i \to N(0, \sigma_\upsilon^2)$ et $0 \leq \phi \leq 1$

– Le taux en protéines mesuré $Z_i$ est, de plus, entaché d'une erreur de mesure :

$$Z_i = Y_i + \varepsilon_i \text{ Où } \varepsilon_i \to N(0, \sigma_\varepsilon^2) \qquad [5]$$

La règle utilisée pour choisir le débit de gluten $u_i$ à l'instant $i$ sera basée sur le débit à l'instant ($i$-$1$) noté par $u_{i-1}$, le taux en protéines $Z_i$ mesuré à l'instant i et la valeur cible $\tau$. Une règle optimale devra être choisie dans le but d'obtenir un taux de protéines dans la farine traitée Y le plus constant possible. Cette règle pourra être formulé en fonction de $u_i$, $Z_i$ et $\tau$ selon l'équation [6].

$$u_i = f(u_{i-1,} Z_i, \tau) \qquad [6]$$

## 3. Modèle d'intégration proposé

### 3.1. *Architecture du modèle*

Pour l'intégration entre MSP et CAP, nous nous proposons de suivre la démarche de la figure 2.

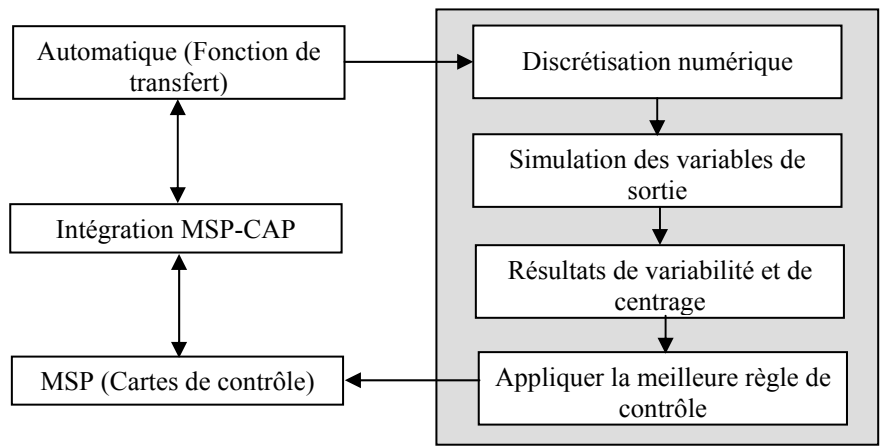

**Figure 2.** *Architecture proposée de l'intégration CAP-MSP*

La première étape de la démarche proposée, consiste à établir les fonctions de transfert des composantes du procédé de fabrication : ajout de gluten, mesure et procédé de mélange. Nous obtenons ainsi les équations différentielles de chaque variable du système.

La deuxième étape consiste à discrétiser les équations différentielles issues de la première étape, c'est-à-dire remplacer les opérateurs continus par ceux qui sont leurs approchés discrets. Il est à noter que dans ce présent travail, nous nous limitons à la règle de contrôle.

Une fois que le modèle de la règle de contrôle est choisit, nous simulons l'évolution de *Z* en fonction du temps. La règle qui donne les meilleurs résultas en terme de la plus petite variabilité et en terme de centrage sera considérée comme meilleure règle de contrôle à adapter.

**3.2.** *Fonction de transfert globale du procédé*

La figure 3 présente la modélisation du procédé de farine selon un système asservi. Le modèle d'intégration proposé consiste à appliquer les concepts d'automatique et de mettre en œuvre le système asservi du modèle représenté par la figure 3. L'équation de transfert globale du procédé, en appliquant l'équation classique des systèmes asservis, s'écrit selon l'équation [7].

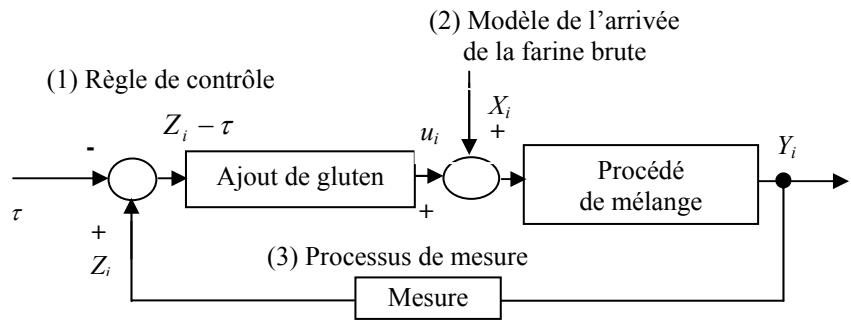

**Figure 3.** *Modélisation du procédé de mélange à l'aide des fonctions de transfert*

$$Y = \tau + \frac{G_2}{1 + G_1 G_2 H} X \qquad [7]$$

Avec :

– $G_2$ désigne la fonction de transfert du mélange. On a $G_2 = 1$ car il constitue simplement un procédé de mélange.

– $G_1$ désigne la fonction de transfert du mécanisme d'ajout de la farine. On a $G_1 = \frac{1}{1+ks}$ car il représente la fonction de transfert d'un système de premier ordre

– H désigne la fonction de transfert de l'appareil de mesure. Dans la littérature de l'automatique des systèmes asservis (Ostertag, 2004), $H = \beta$ avec β est une constante.

On obtient donc :

$$Y = \tau + \cfrac{1}{1+\cfrac{1}{1+ks}\beta} X = \tau + \frac{1+ks}{(1+ks)+\beta} X \qquad [8]$$

En automatique, cette fonction de transfert nous permet d'étudier le comportement du procédé et en particulier l'étude de stabilité. Ceci fera l'objet de travaux de recherche encours. En fournissant cette première vue d'ensemble complète de la fonction de transfert, de futurs développements de recherche et perfectionnement seront plus faciles à entreprendre.

### 3.3. *Choix de la règle de contrôle*

La règle de contrôle est adoptée par le décideur. En effet, elle constitue une variable de décision. Une règle de contrôle est caractérisée par son efficacité, c'est-à-dire par la variabilité et le centrage de la variable de sortie par rapport à la cible.

3.3.1. *La règle de contrôle proposée par (Fearn et Maris, 1990)*

La règle qui a été proposé par (Fearn et Maris, 1990) consiste à *calculer* la différence entre le taux $Z_i$ mesuré et la valeur cible $\tau$ et de corriger le débit u pour tenter d'annuler cette différence :

$$u_i = u_{i-1} - (Z_i - \tau) \qquad [9]$$

On note d(t) le signal d'entrée, u(t) le signal de sortie, D(s) la transformé de Laplace de d(t) et U(s) la transformé de Laplace de u(t). Le modèle de l'ajout de gluten peut s'écrire selon la fonction de transfert présentée dans l'équation (10).

$$F(s) = \frac{U(s)}{D(s)} = -\frac{1}{s} \qquad [10]$$

En utilisant les concepts d'automatique, l'équation différentielle réagissant la règle de contrôle est donnée dans l'équation [11].

$$\frac{du(t)}{dt} = -d(t) = -(Z(t) - \tau) \qquad [11]$$

En discrétisant l'équation [11], on obtient :

$$u_i - u_{i-1} = -(Z_i - \tau) \qquad [12]$$

3.3.2. *La règle de contrôle proposée*

La règle proposée se base sur un système reconnu en automatique par un système de premier ordre. L'avantage de cette règle réside dans sa facilité de mise en œuvre. La fonction de transfert d'un système de premier ordre est donnée par l'équation [13]

$$F(s) = \frac{U(s)}{D(s)} = \frac{1}{1+ks} \qquad [13]$$

En appliquant la transformé inverse de Laplace, l'équation différentielle du système du premier ordre est donnée par l'équation [14]

$$k\frac{du}{dt} + u(t) = d(t) \qquad [14]$$

Après discrétisation à gauche, nous obtenons l'équation [15]

$$k(u_i - u_{i-1}) + u_{i-1} = d_i = Z_i - \tau \qquad [15]$$

L'une des solutions possibles est donnée par l'équation [16]

$$u_i - u_0 = \lambda_1(u_{i-1} - u_0) - \lambda_2(Z_i - \tau) \qquad [16]$$

Où $\lambda_1$ et $\lambda_2$ deux constantes à fixer par le décideur. Il est à remarquer que nous pouvons retrouver l'expression de la règle de (Fearn et Maris, 1990) dans l'équation [9], en prenant $\lambda_1 = \lambda_2 = 1$.

**3.4.** *Méthodologie d'intégration*

La procédure d'intégration comporte quatre étapes :

***Étape 1*** *:* Développer un modèle qui permet de décrire la dynamique de la (ou des) variable(s) d'entrées du système et les perturbations qu'elles subissent. Ceci peut par exemple consister à identifier et à estimer un modèle de type AR (p) pour la variable donnée. Pour l'exemple sujet de cette étude, le taux de protéine dans la farine brute (Xi) suit, par hypothèse, un modèle de type AR (1) et nous supposons disposer de données passées pour en estimer les paramètres.

***Étape 2* :** Établir une carte de contrôle pour la variable d'entrée afin de maîtriser la variabilité. Quand une carte de contrôle est en alarme, il s'agira de rechercher les causes. Pour l'exemple étudié, il s'agit de mettre en place une carte de contrôle des résidus à valeurs individuelles étendue mobile, notée en littérature par (X, Rm), pour le taux de protéine dans la farine brute (*Xi*). L'objectif de cette carte de contrôle est de tester l'hypothèse de la normalité des résidus. Il est à remarquer qu'il est nécessaire de vérifier que les coefficients du modèle d'arrivée de la farine brute restent constants au cours du temps.

*Étape 3 :* Choisir une règle de contrôle en accord avec le modèle et la structure du procédé (variables de contrôle disponible, coût des ajustements, etc.).

*Étape 4 :* Implémenter la règle de contrôle sur le procédé en parallèle avec une carte de suivi : l'objectif est de vérifier simultanément la bonne marche du procédé et du système de contrôle. La carte de suivi devra signaler quand le procédé ou le régulateur n'est plus valable. Quand une carte de suivi est en alarme, il s'agira de rechercher les causes dans le procédé (le taux de protéine dans la farine brute et\ou la capabilité des moyens de mesure) et de tenter de les éliminer. Si aucune cause n'est trouvée dans le procédé, on remettra en cause la règle de contrôle et/ou le modèle choisit de l'arrivée de la farine brute.

**3.5.** *Etablissement des cartes de contrôle*

Pour l'établissement de la carte de contrôle, nous avons simulé les données relatives à l'arrivée de la farine brute selon le modèle AR (1) qui sont présentées dans le tableau 1.

Avec : $X_i - \mu = \phi(X_{i-1} - \mu) + \upsilon_i$ ; $X_0 = \mu = 10$ ; $\phi = 0{,}7$ et $\varepsilon_i \rightarrow N(0.1)$

Les deux hypothèses de Shewhart pour l'établissement d'une carte de contrôle sont l'indépendance et la normalité des variables à mesurer. C'est pourquoi, une carte de contrôle associée à une série temporelle nécessite un traitement statistique spécifique pour valider ces deux hypothèses.

| i | Xi | i | Xi | i | Xi | i | Xi | i | Xi |
|---|---|---|---|---|---|---|---|---|---|
| 1 | 10,00 | 11 | 8,27 | 21 | 9,78 | 31 | 9,34 | 41 | 7,31 |
| 2 | 8,66 | 12 | 8,25 | 22 | 10,72 | 32 | 10,74 | 42 | 7,59 |
| 3 | 9,43 | 13 | 9,00 | 23 | 11,95 | 33 | 10,42 | 43 | 7,51 |
| 4 | 9,19 | 14 | 7,90 | 24 | 10,77 | 34 | 11,45 | 44 | 8,15 |
| 5 | 10,67 | 15 | 8,45 | 25 | 10,25 | 35 | 10,95 | 45 | 7,64 |
| 6 | 10,17 | 16 | 7,67 | 26 | 10,18 | 36 | 10,15 | 46 | 9,27 |
| 7 | 8,85 | 17 | 5,00 | 27 | 8,86 | 37 | 9,42 | 47 | 10,58 |
| 8 | 8,79 | 18 | 6,11 | 28 | 8,11 | 38 | 7,73 | 48 | 11,57 |
| 9 | 8,82 | 19 | 7,47 | 29 | 7,97 | 39 | 6,81 | 49 | 10,23 |
| 10 | 9,48 | 20 | 9,31 | 30 | 7,95 | 40 | 7,30 | 50 | 11,95 |

**Tableau 1.** *Taux de gluten de la farine brute AR (1)*

Avant de commencer l'établissement de la carte de contrôle des résidus, nous avons vérifié la stationnarité et l'auto corrélation de la série du tableau 1. Pour cela, nous avons appliqué le Test de Dickey_Fuller pour l'étude de stabilité. Puis nous avons tracé le corrélogramme correspondant, et ceci en utilisant le logiciel *Eviewes*. Ensuite, nous avons estimé les paramètres du modèle AR (1) comme l'indique la figure 4.

L'examen de la figure 4 montre que le paramètre $\hat{\phi}$ est estimé par $\hat{\phi} = 0{,}669$ et la constante C est estimée par $\hat{C} = 8{,}856$. Le résidu qu'on le note par ($x_i$) du modèle est donné par l'équation [16].

$$x_i = X_i - \hat{\phi}(X_{i-1}) - \hat{C} \qquad [16]$$

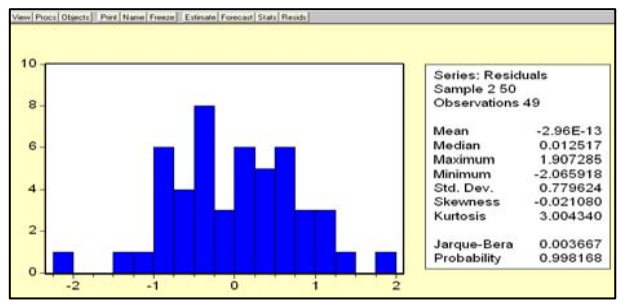

**Figure 4.** *Estimation des coefficients du modèle AR (1)*

Enfin, nous avons vérifié la normalité des résidus par le test de Jarque-Bera à l'aide du logiciel *Eviews,* comme l'indique la figure 5.

**Figure 5.** *Test de normalité des résidus*

Pour étudier la stabilité de l'arrivée de la farine brute sous la forme d'un modèle AR (1), nous proposons de tracer la carte des étendues mobile (x, $R_M$) pour la détection des grandes déréglages, puisque on a validé les hypothèses de la normalité et de l'absence d'auto corrélation.

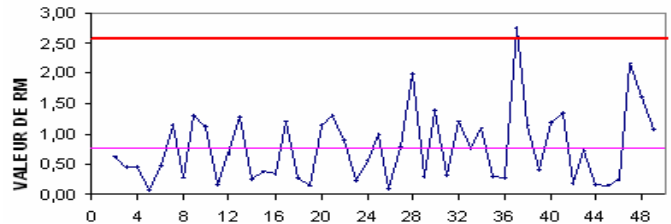

**Figure 6.** *Carte des étendues mobiles des résidus*

L'examen de la figure 6 montre un point aberrant. C'est une indication sur la présence des causes assignables qui va perturber la variabilité du procédé. A ce niveau, il serait indispensable de consulter les fiches de contrôle à la réception (le journal de bord) pour identifier les causes possibles de ces anomalies.

## 4. Simulation et analyse des résultats

Cette section à pour but de choisir la règle de contrôle qui donne les meilleures performances du procédé. Les figures 7 et 8 présentent, respectivement, les résultats de simulation du variable de sortie $Z_i$, pour la règle de contrôle proposée par (Fearn et Maris, 1990) et pour la règle qu'on a proposée. Les simulations sont réalisées pour les paramètres suivants :

$$X_0 = \mu = 10, \ u_0 = 6, \phi = 0{,}7, \tau = 16, \sigma_\nu^2 = 0{,}5 \ et \ \sigma_\varepsilon^2 = 0{,}5$$

Nous avons réalisés 30 réplications pour chaque règle de contrôle. Les résultats présentés dans les figures 7 et 8 constituent une moyenne des différentes réplications. Pour chaque règle de contrôle, les résultats de chaque réplication, constituent des variables aléatoires indépendantes (générateur de nombres pseudo aléatoire) et suivent la même loi (simulations faites sur le même modèle). En appliquant le théorème de central limite, on peut simplement déduire que la moyenne des 30 réplications suit une loi normale qui sera alors caractérisée par sa moyenne E et son écart type (ou variance V).

### 4.1. *La règle de contrôle proposée par (Fearn et Marris, 1990)*

La figure 7 représente l'évolution de la variable de sortie $Z_i$ pour le cas où la règle de contrôle de (Fearn et Maris, 1990) selon l'équation (9).

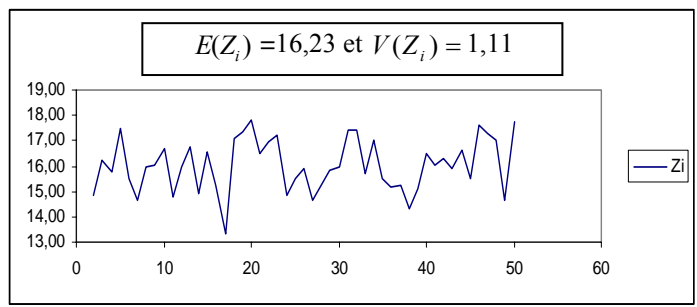

**Figure 7.** *Evolution temporelle de Z pour la règle de (Fearn et Marris, 1990)*

### 4.2. *La règle de contrôle proposée*

La figure 8 représente l'évolution de la variable de sortie $Z_i$ pour le cas de la règle de contrôle que nous avons proposé :

$$u_i - u_0 = \lambda_1 (u_{i-1} - u_0) - \lambda_2 (Z_i - \tau) \text{ avec } \lambda_1 = \lambda_2 = 0{,}5$$

Pour la première règle, le taux de protéines de la farine traitée $Z_i$ n'est pas centré et assez variable. En effet, l'espérance ($E(Z_i)$=16,23) est assez loin de la cible et la variance ($V(Z_i) = 1{,}11$). Pour la seconde règle que nous avons proposé, $Z_i$ est plus centrée ($E(Z_i)$=16,01) et possède moins de variabilité ($V(Z_i) = 0{,}87$).

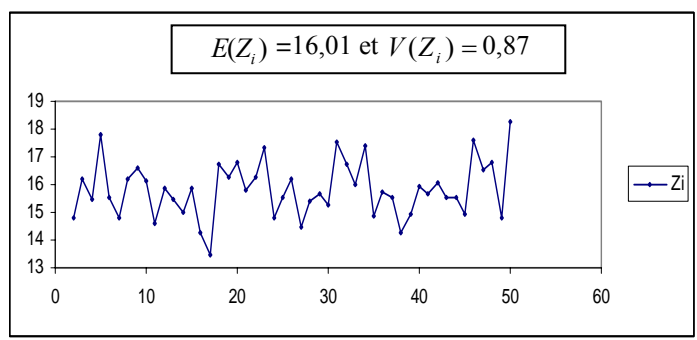

**Figure 8.** *Evolution temporelle de Z pour la règle de contrôle proposée*

Ainsi, la première règle de contrôle ne permet pas de récupérer la dérive. Le taux de protéine de la farine traitée s'éloigne progressivement de la valeur cible. Mais, à travers notre proposition qui consiste à ajouter au régulateur une composante différentielle, on peut centrer le procédé et réduire les variations qui y manifestent.

## 5. Conclusion

Ce travail porte sur la proposition d'un modèle d'intégration du CAP dans la MSP. Cette intégration se base sur l'établissement d'une carte de contrôle des résidus à l'entrée du système et sur le choix de la règle de contrôle la plus adéquate. Dans ce travail, nous avons proposé une règle de contrôle qui se base sur un système de premier ordre. Cette règle a réduit la variabilité comparant à celle proposée en littérature.

Nous avons appliqué le modèle développé à un cas particulier d'un procédé de production continu. Des perspectives de ce travail consistent d'une part, à rechercher la configuration optimale de la règle de contrôle proposée, c'est-à-dire les coefficients $\lambda_1$ et $\lambda_2$. D'autre part, à valider la structure de ce modèle pour des modèles statistiques de l'arrivée de la matière brute plus générale de type ARMA, ARIMA, etc.

## 6. Bibliographie